\documentclass[12pt,preprint]{aastex}
\usepackage{natbib}

\begin{document}

\title{\bf NSV 11154 Is a New R Coronae Borealis Star}
\author{Nutsinee Kijbunchoo\altaffilmark{1}, Geoffrey C. Clayton\altaffilmark{1}, Timothy C. Vieux\altaffilmark{1}, N. Dickerman\altaffilmark{1}, T. C. Hillwig\altaffilmark{2}, D. L. Welch\altaffilmark{3}, Ashley Pagnotta\altaffilmark{1}, Sumin Tang\altaffilmark{4}, J. E. Grindlay\altaffilmark{4}, and A. Henden\altaffilmark{5}}

\altaffiltext{1}{Department of Physics \& Astronomy, Louisiana State University, Baton Rouge, LA 70803 USA; nkijbu1@tigers.lsu.edu, gclayton@fenway.phys.lsu.edu, tvieux1@tigers.lsu.edu,\\ pagnotta@phys.lsu.edu}
\altaffiltext{2}{Department of Physics and Astronomy, Valparaiso University, Valparaiso, IN 46383; todd.hillwig@valpo.edu}
\altaffiltext{3}{Department of Physics and Astronomy, McMaster University, 1280 Main Street West, Hamilton, ON L8S 4M1 Canada; welch@physics.mcmaster.ca}
\altaffiltext{4}{Harvard-Smithsonian Center for Astrophysics, 60 Garden Street, Cambridge, MA 02138; stang@cfa.harvard.edu, josh@head.cfa.harvard.edu}
\altaffiltext{5}{American Association of Variable Star Observers, 49 Bay State Rd., Cambridge, MA 02138; arne@aavso.org}


\begin{abstract} NSV 11154 has been confirmed as a new member of the rare hydrogen-deficient R Coronae Borealis (RCB) stars based on new photometric and spectroscopic data. Using new photometry, as well as archival plates from the Harvard archive, we have constructed the historical lightcurve of NSV 11154 from 1896 to the present. The lightcurve shows the sudden, deep, irregularly spaced declines characteristic of RCB stars. The visible spectrum is typical of a cool (T$_{eff}$ $\lesssim$ 5000 K) RCB star showing no hydrogen lines, strong $C_2$ Swan bands, and no evidence of  $^{13}C$. In addition, the star shows small pulsations typical of an RCB star, and an infrared excess due to circumstellar dust with a temperature of $\sim$800 K. The distance to NSV 11154 is estimated to be $\sim$14.5 kpc. 
RCB stars are very rare in the Galaxy so each additional star is important to population studies leading to a better understanding the origins of these mysterious stars. Among the known sample of RCB stars, NSV 11154 is unusual in that it
lies well above the Galactic plane (5 kpc) and away from the Galactic Center
which suggests that its parent population is neither thick disk nor bulge.

\end{abstract}

\section{\bf Introduction} 

The R Coronae Borealis (RCB) stars represent an extremely rare class of variable stars  \citep{1996PASP..108..225C}. They are cool supergiants, which are carbon-rich and hydrogen deficient. Their defining characteristic is large irregular declines in brightness of up to 8 mag caused by the formation of carbon dust. Two scenarios have been suggested which attempt to clarify the origins of the RCB stars, the double degenerate (DD), and the final helium-shell flash (FF). The DD model suggests that RCB stars are formed by the merger of a CO- and a He-white dwarf (WD), and the FF model involves stellar expansion after a helium-shell flash. The high $^{18}$O/$ ^{16}$O ratios found in RCB stars favor the DD model. However a few RCB stars show Li in their spectra which may instead favor the FF model \citep{1996ApJ...456..750I,1996PASP..108..225C} . 

There could be as many as 3000 RCB stars in the Galaxy based on the numbers found in the Large Magellanic Cloud (LMC), but only 55 have been discovered in the Galaxy so far \citep{1996PASP..108..225C,2001ApJ...554..298A,2005AJ....130.2293Z,2008A&A...481..673T}. 

About forty-five years ago, \citet{1966AN....289..139H} discovered that NSV 11154 was a variable star (S 9323 Lyr) and suggested that NSV 11154 is a short periodic variable.
NSV 11154 was also found to be variable in the ROTSE-I survey with an amplitude of 0.4 mag and  it was suggested that NSV 11154 is a long period variable
\citep{2000AJ....119.1901A,2001IBVS.5134....1W}.
\citet{2009IBVS.5890....1H} 
examined 562 plates, obtained at Sonnenberg Observatory during 1964--1996, and found irregular brightness variations between 13.0 and 17.2 mag. On the basis of the lightcurve, they suggested that NSV 11154 may be an RCB star. 
In this article, we use newly acquired photometry and spectroscopy to attempt to confirm Haussler et al.'s suggestion that NSV 11154 is indeed an RCB star.

\section{\bf OBSERVATIONS AND DATA REDUCTION}

The UCAC3 coordinates of NSV 11154 are
$\alpha$(2000) $18^{h}~37^{m}~51\fs254$ $\delta(2000) +47\deg~23\arcmin~23\farcs45$ \citep{2010AJ....139.2184Z}. The field of NSV 11154 is shown in Figure 1.

The Sonnenberg Plate data, taken by a 40 cm astrograph on blue-sensitive photographic plates \citep{2009IBVS.5890....1H}, were downloaded for use in this study.   
The blue photographic magnitudes were transformed to Johnson V using,  
V - $m_{pg}$ = 0.17 - 1.09(B - V) \citep{1961ApJ...133..869A}. The transformed data are plotted in Figures 2 and 4.
There were 250 additional plates of the NSV 11154 field dating as  far back as 1896, available from the Harvard College Observatory plate archive. 
These plates have been scanned and photometry was done on the stars as part of the Digital Access to a Sky Century at Harvard (DASCH) program \citep{2009ASPC..410..101G}, from the scanning focused in the Kepler field. Note that
NSV 11154 is a few degrees outside the Kepler field of view, and there
are many more Harvard plates covering this star not scanned yet.
The measured magnitudes in the DASCH database were converted from photographic magnitudes to Kepler Input Catalog (Sloan) g magnitudes 
\citep{2011arXiv1102.0342B}. 
The DASCH data were then converted from g to Johnson V using, V=g\arcmin~+ 0.12 - 0.56(B - V) \citep{1996AJ....111.1748F}. 
The DASCH data used here came from twelve plate series from the archive covering the period 1896-1989, and have average uncertainties of 0.15 mag \citep{2010AJ....140.1062L}. This will be further improved with photometric corrections now being optimized (Tang et al. 2011, in preparation).  The Sonnenberg plates were obtained during 1964-1996 so there is an overlap of roughly twenty years with the DASCH data. 
The DASCH photometry is  listed in Table 1 and is plotted in Figures 2 and 4.

  The ROTSE-I Northern Sky Variability Survey (NSVS) detected NSV 11154 as a variable \citep{2000AJ....119.1901A}. The photometric data were downloaded from the NSVS archive. The ROTSE-I images are unfiltered and
  \citet{2004AJ....127.2436W} suggest that m$_{ROTSE}$ is equivalent to Johnson V, when it is actually very close to Cousins R \citep{2005IBVS.5620....1B}. If the relation between m$_{ROTSE}$ and V$_T$ \citep{2004AJ....127.2436W}  is combined with the equations to convert from V$_T$ to Johnson V\footnote{http://heasarc.nasa.gov/W3Browse/all/tycho2.html},  then the conversion is equivalent to V = m$_{ROTSE}$ + (V-R)$_C$ =  
  m$_{ROTSE}$ + 0.55. The converted ROTSE-I photometry still seems to have a systematic shift from the actual Johnson V photometry (see below) of $\sim$0.2 mag.
  The ROTSE-I photometry is listed in Table 2, and is plotted in Figures 2 and 4.
 
New BVR$_C$I$_C$ photometry has been obtained with the AAVSO Sonoita Research Observatory (SRO) between 2010 October and 2011 June. The images were obtained with the 35cm C14 OTA (SRO35) and the 50cm f/4 Newtonian (SRO50). The images were flat-fielded and dark subtracted. 
Aperture photometry was done using DAOPHOT in IRAF. 
The instrumental BV magnitudes were transformed to standard magnitudes using a photometric BV sequence of the field (5169jnc) provided by the AAVSO. In particular, two stars, 
2MASS 18374206+4723474 and 18374814+4724267, were used. These stars are both in the Kepler Input Catalog \citep{2011arXiv1102.0342B}. Their Sloan gri magnitudes were transformed to Cousins R and I \citep{1996AJ....111.1748F} and then were used to transform the SRO RI instrumental magnitudes of NSV 11154 to standard R$_C$I$_C$ magnitudes. 
The uncertainties are $\sim$0.01-0.02 mag. 
The BVR$_C$I$_C$ photometry is listed in Table 3 and plotted in Figures 3 and 4.
In addition, there is also photometry from 2MASS, AKARI, and IRAS for NSV 11154. These data are tabulated in Table 4. The 2MASS data were obtained during a gap between the Sonnenberg and ROTSE-I photometry but the star appears to be at or near maximum light.

Spectra of NSV 11154 were obtained on 2009 August 18 using the 4m telescope at Kitt Peak National Observatory (KPNO) with the RC spectrograph using the BL380 grating which has a resolution of 0.9 \AA.  Three consecutive 600s spectra were summed.  The spectra were not  flux calibrated.  The wavelength calibration has an rms uncertainty of about 0.02 \AA. 
The summed spectrum is plotted in Figure 5 along with the spectrum of a similar RCB star HV 5637 \citep{2001ApJ...554..298A}.

\section{DISCUSSION}

The historical NSV 11154 lightcurve, seen in Figure 4 from 1896 to 2011, is fragmentary, but several deep declines are apparent. The last decline detected was in 1996, but the coverage since then has been spotty. No declines are seen in the recent ROTSE-I or SRO photometry.
Although the lightcurve data are sparse, NSV 11154 seems to be an active RCB star having frequent declines. 
There are at least 13 epochs where NSV 11154 is seen 2 mag or more below maximum light. These are listed in Table 5. 
The characteristic time between declines in RCB stars is typically about 1000 days, but there is a wide range in activity among the RCB stars \citep{1986ASSL..128..151F,1996AcA....46..325J}. 
From the ROTSE-I and SRO lightcurves it can be seen that NSV 11154 pulsates with period of $\sim$50 days between 1999-2000, and $\sim$40 days in 2010-2011. The pulsations have an amplitude of $\sim$0.4 mag.

Figure 5 shows the visible spectrum of NSV 11154 compared to the LMC RCB star HV 5637. 
The spectrum of NSV 11154 is typical of a cool ($\lesssim$5000 K) RCB star with strong CN and 
C$_2$ absorption bands \citep{1996PASP..108..225C,2001ApJ...554..298A}. 
There is no sign of either H$\beta$ or H$\gamma$ indicating extreme hydrogen deficiency. The CH band at 4300 \AA\ is also absent. In addition, the $^{12}$C$^{13}$C band at 4744 \AA\ is weak or absent, while the  $^{12}$C$^{12}$C band at 4737 \AA\ is very strong indicating a high $^{12}$C to $^{13}$C ratio. This is typical of most RCB stars. 
 NSV 11154 lies well out of the Galactic plane at b$^{II}$ = +21\fdg8, and so the estimated foreground extinction is quite small, E(B-V) = 0.07 mag \citep{1998ApJ...500..525S}. NSV 11154 has an observed
(B-V)=1.1 mag which is consistent with it being $\lesssim$5000 K and lightly reddened \citep{1990MNRAS.247...91L}. 

The measured colors of NSV 11154, B-V = 1.1 and V-I = 1.2, assuming little or no reddening, are consistent with an absolute magnitude of M$_V$= -4 mag \citep{2001ApJ...554..298A,
2009A&A...501..985T}. Then, if the foreground extinction is A$_V\sim$0.2 mag, the distance to NSV 11154 is 
14.5$ \pm$ 1.5 kpc. This also implies that NSV 11154 is 5.4 kpc above the Galactic plane. Most RCB stars are within 2 kpc of the plane \citep{2005AJ....130.2293Z}. Only two other RCB stars in the Galaxy, UX Ant and U Aqr, are as far above or below the plane as NSV 11154. Also, most RCB stars are seen toward the Galactic center, but NSV 11154 has l$^{II}$ = 76\arcdeg.
NSV 11154 lies well away from the extended body of the Sagittarius Dwarf Galaxy 
\citep{2003ApJ...599.1082M}.

Using the photometry in Table 4, an SED for NSV 11154 has been plotted in Figure 6. This SED can be fit very well by two blackbodies with temperatures of 4500 and 800 K corresponding to the RCB star itself and its circumstellar dust, respectively.
The RCB dust shells typically have temperatures range from 600 K to 900 K  \citep{1985A&A...152...58W}.

\section{Summary}

The suggestion of \citet{2009IBVS.5890....1H}, on the basis of the lightcurve, that NSV 11154 is an RCB star was correct. 
The spectrum and colors of NSV 11154 show it to be a cool ($\lesssim$5000 K) RCB star. 
Archival photometry as well as new BVR$_C$I$_C$ photometry have been collected giving a historical lightcurve from 1896 to the present. The new photometry shows that NSV 11154 has a semi-regular pulsation period of 40-50 d with an amplitude of 0.4 mag. Although the lightcurve is fragmentary, NSV 11154 has had a number of deep declines showing it to be an active RCB star.
The star also displays a significant IR excess indicating the presence of dust with T$\sim$800 K, which is typical of RCB stars. 

Only 55 other RCB stars are known in the Galaxy, so each addition is important to population studies, which will help us to better understand the origins of these mysterious stars. NSV 11154 lies well above the Galactic plane quite different from most other RCB stars which seem to fit into an old disk or bulge population. 
This may favor the final flash model for this star since the higher stellar density of the Galactic center region is more conducive to formation of RCB stars by the double degenerate scenario.
 Despite being very rare, RCB stars may be a key to understanding the late stages of stellar evolution. Their measured isotopic abundances imply that many RCB stars are produced by the mergers of double degenerate white dwarfs, which may be the low-mass counterparts of the more massive mergers thought to produce type Ia supernovae. Therefore, knowing the population of RCB stars in the Galaxy will help determine the frequency of these white dwarf mergers.

\acknowledgments
This paper used data from Digital Access to a Sky Century at Harvard (DASCH) supported by NSF grants AST-0407380 and AST-0909073, as well as by the {\it Cornel and Cynthia K. Sarosdy Fund}.

\bibliographystyle{/Users/gclayton/projects/latexstuff/apj}

\bibliography{/Users/gclayton/projects/latexstuff/everything2}

\clearpage

\begin{figure}
  \centering
   \includegraphics[width=5in]{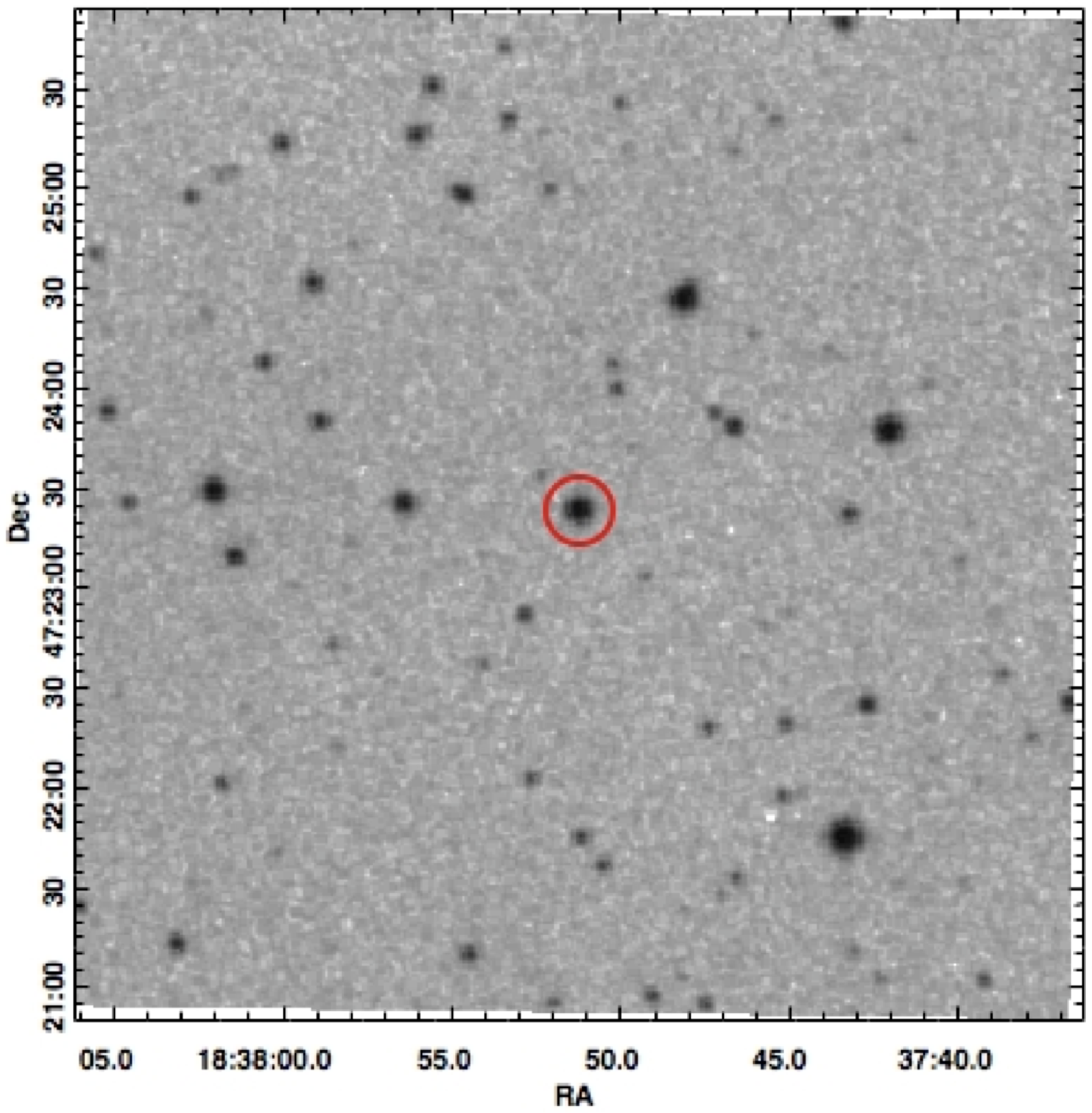} 
   \caption{Identification chart for NSV 11154 from a POSS-II N plate obtained 1992 June 18. The field is 5\arcmin x 5\arcmin. The coordinates are J2000.0. North is up and East is to the left.  }
   \label{fig:1}
   \end{figure}

\begin{figure}[t] 
  \centering
   \includegraphics[width=6in]{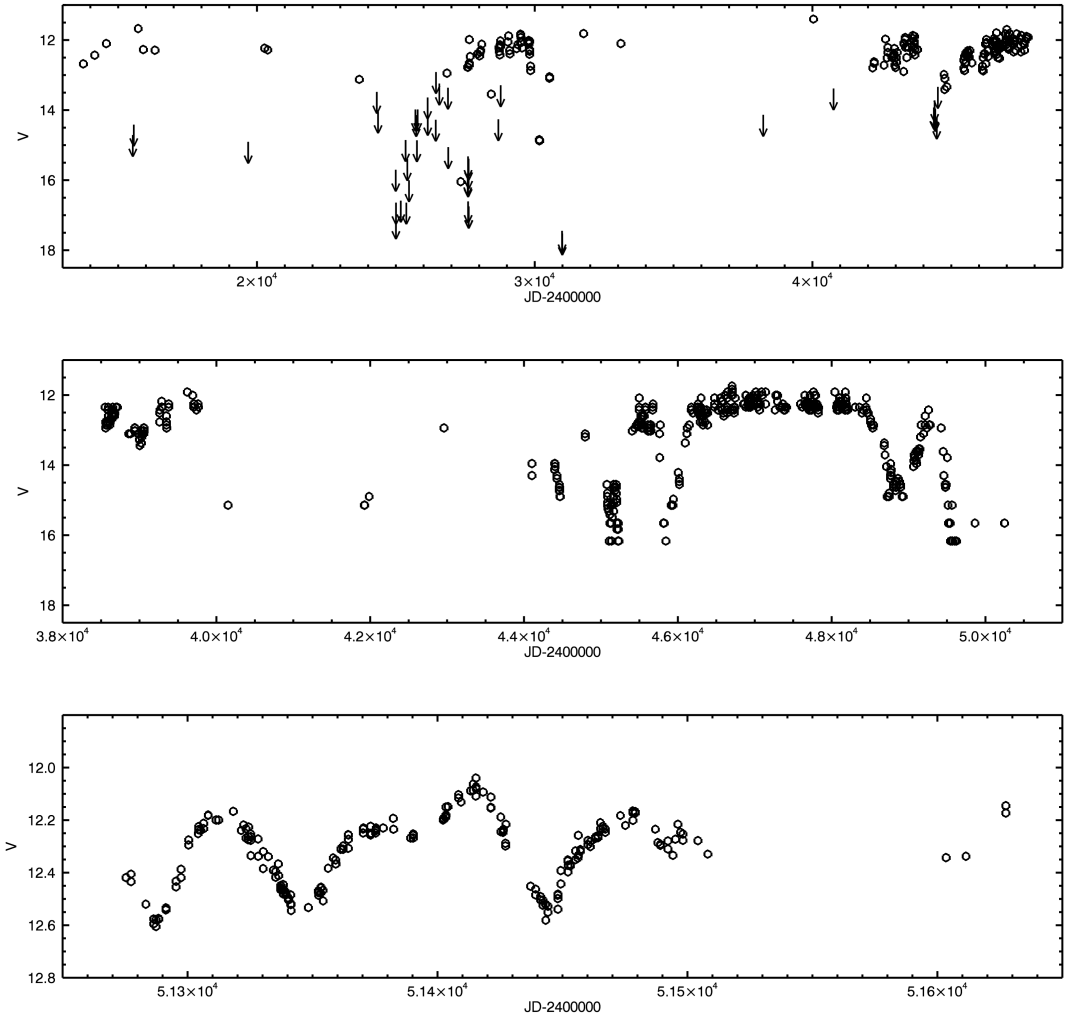} 
   \caption{V-band lightcurves of NSV 11154. Top: DASCH plate photometry and upper limits (arrows). Middle: Sonneberg plate photometry . Bottom: ROTSE-I photometry. }
   \label{fig:1}
   \end{figure}
   
   \begin{figure}[t] 
  \centering
   \includegraphics[width=6in]{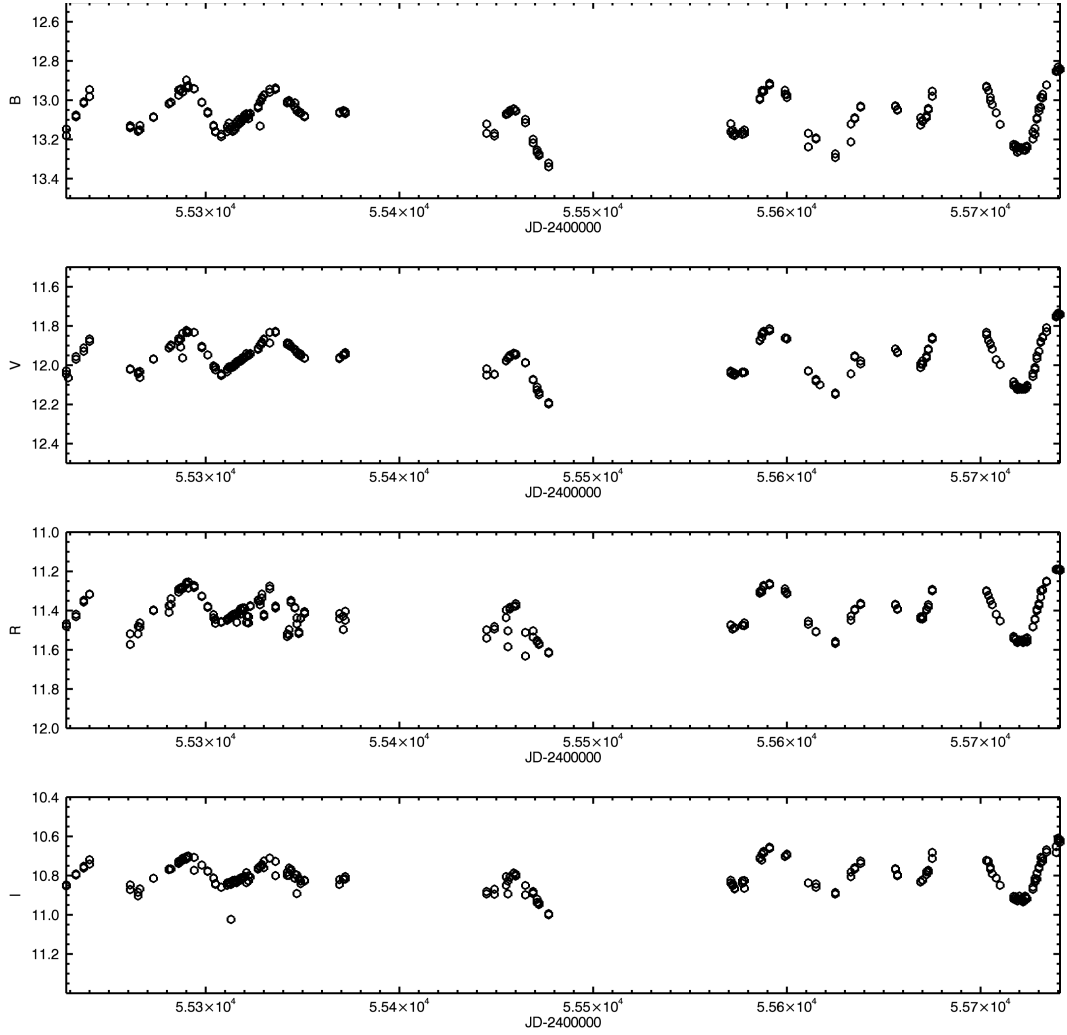} 
   \caption{SRO lightcurves of NSV 11154. The four panels from the top are B, V, R$_C$ and I$_C$ plotted vs JD.}
   \label{fig:1}
   \end{figure}
   
   \begin{figure}[t] 
  \centering
   \includegraphics[width=5in]{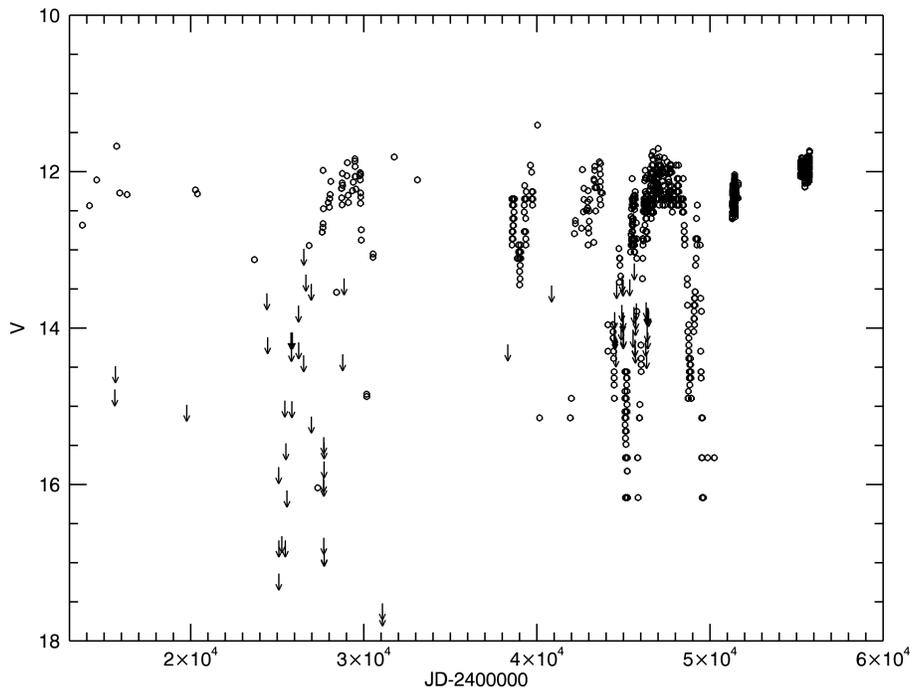} 
   \caption{Historical lightcurve of NSV 11154 from 1896-2011. Symbols are the same as in Figure 2. Several deep declines are detected.}
   \label{fig:1}
   \end{figure}

     \begin{figure}[t] 
  \centering
   \includegraphics[width=5in]{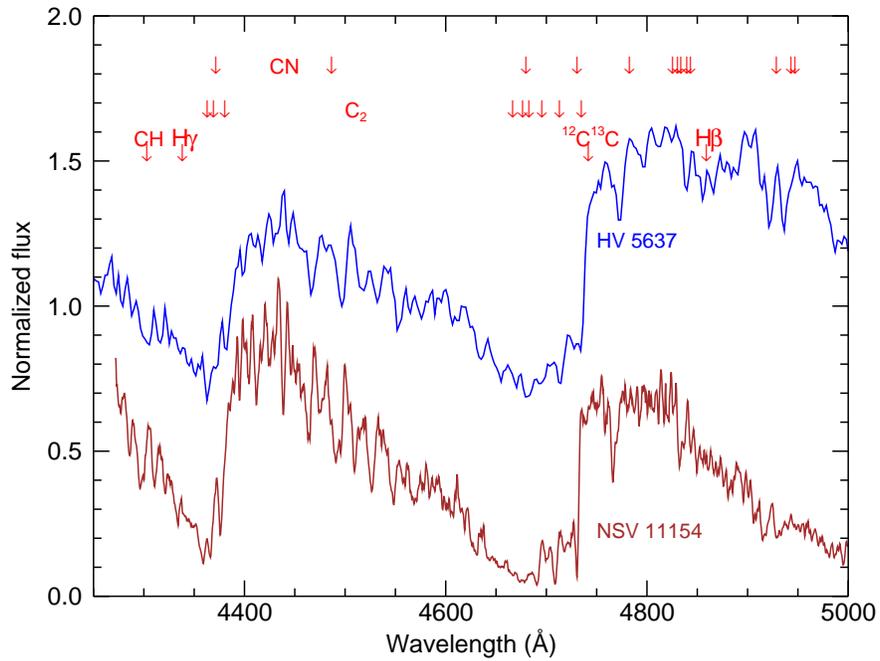} 
   \caption{The spectrum of NSV 11154, together with the spectrum of HV 5637 a known cool RCB star, for comparison. Note the absence of hydrogen, the strong CN and C$_2$ bands, and the absence of $^{13}$C. The spectrum of NSV 11154 is typical for a cool ($\lesssim$5000 K) RCB star. }
   \label{fig:2}
   \end{figure}
   
   \begin{figure}[h] 
  \centering
   \includegraphics[width=5in]{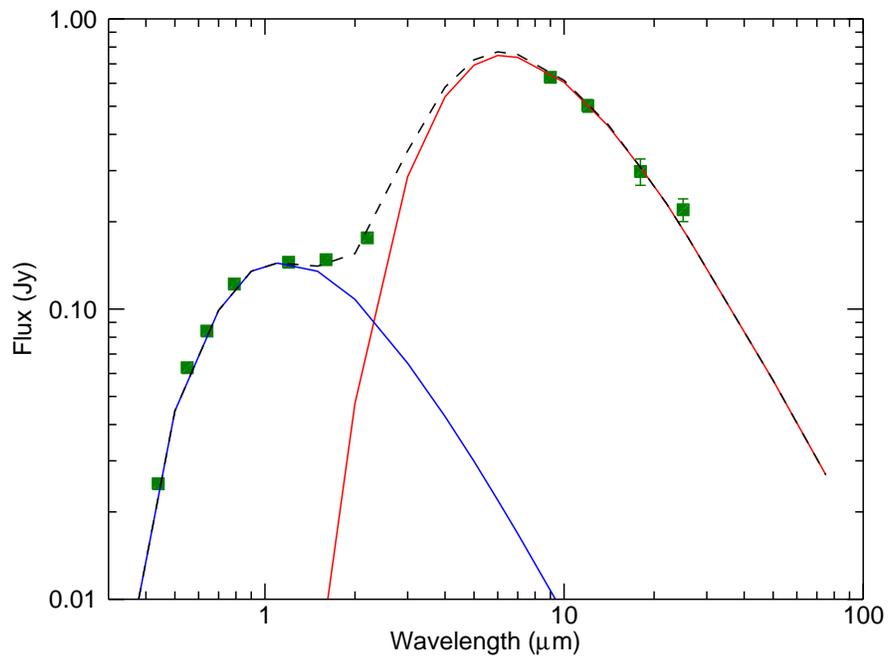} 
   \caption{The SED of NSV 11154 showing the photometry listed in Table 4 (filled squares). Two blackbodies representing the star and the dust shell, 4500 K (blue line) and 800 K (red line), have been fit to the data. The dashed line is the sum of the two blackbodies. }
\label{fig:3}
   \end{figure}

\clearpage


\begin{deluxetable}{cc}  
\tablecolumns{2}
\tablecaption{DASCH Photometry}
\tablehead{   
\colhead{JD} &
\colhead{V} 
}
\startdata
2413740.74	&	12.7	\\
2414154.68	&	12.4	\\
2414569.61	&	12.1	\\
2415150.74	&	$>$14.8	\\
2415186.80	&	$>$14.5	\\
2415717.46	&	11.7	\\
2415901.67	&	12.3	\\
2416323.76	&	12.3	\\
2419306.50	&	$>$15.0	\\
2420270.83	&	12.2	\\
2420385.57	&	12.3	\\
2423680.54	&	13.1	\\
2423937.79	&	$>$13.6	\\
2423981.69	&	$>$14.2	\\
2424618.50	&	$>$15.8	\\
2424624.50	&	$>$17.2	\\
2424626.50	&	$>$16.8	\\
2424800.50	&	$>$16.7	\\
2424971.50	&	$>$15.0	\\
2424999.50	&	$>$16.8	\\
2425035.50	&	$>$15.5	\\
2425096.50	&	$>$16.1	\\
2425327.87	&	$>$14.1	\\
2425354.82	&	$>$14.3	\\
2425380.75	&	$>$15.0	\\
2425408.72	&	$>$14.1	\\
2425765.68	&	$>$13.8	\\
2425771.69	&	$>$14.2	\\
2426059.88	&	$>$14.4	\\
2426067.85	&	$>$13.0	\\
2426188.62	&	$>$13.4	\\
2426498.70	&	$>$13.5	\\
2426507.78	&	$>$15.2	\\
2426837.79	&	12.9	\\
2427210.78	&	$>$15.9	\\
2427222.74	&	$>$15.4	\\
2427225.75	&	$>$16.0	\\
2427225.80	&	$>$16.7	\\
2427241.75	&	$>$15.5	\\
2427241.79	&	$>$15.7	\\
2427246.74	&	$>$16.9	\\
2427246.79	&	$>$16.9	\\
2427334.63	&	16.0	\\
2427581.70	&	12.8	\\
2427597.75	&	12.8	\\
2427640.62	&	12.7	\\
2427640.76	&	12.7	\\
2427642.63	&	12.0	\\
2427668.61	&	12.5	\\
2427938.73	&	12.4	\\
2428011.70	&	12.5	\\
2428023.60	&	12.3	\\
2428052.56	&	12.3	\\
2428089.52	&	12.1	\\
2428314.80	&	$>$14.4	\\
2428397.59	&	$>$13.4	\\
2428428.57	&	13.5	\\
2428701.76	&	12.2	\\
2428717.69	&	12.3	\\
2428721.68	&	12.2	\\
2428743.66	&	12.4	\\
2428749.64	&	12.1	\\
2428760.59	&	12.0	\\
2428774.63	&	12.2	\\
2429027.81	&	12.1	\\
2429052.78	&	11.9	\\
2429090.71	&	12.3	\\
2429104.60	&	12.4	\\
2429344.87	&	12.2	\\
2429394.81	&	12.1	\\
2429483.67	&	11.9	\\
2429485.66	&	11.8	\\
2429488.61	&	12.1	\\
2429499.61	&	11.9	\\
2429521.55	&	12.1	\\
2429547.52	&	12.2	\\
2429759.81	&	12.1	\\
2429787.76	&	12.0	\\
2429788.75	&	12.1	\\
2429810.67	&	12.4	\\
2429810.74	&	12.0	\\
2429812.64	&	12.3	\\
2429813.63	&	12.3	\\
2429846.60	&	12.7	\\
2429847.64	&	12.9	\\
2430163.68	&	14.8	\\
2430167.78	&	14.9	\\
2430529.77	&	13.1	\\
2430531.72	&	13.1	\\
2430608.61	&	$>$17.6	\\
2430616.59	&	$>$17.6	\\
2431759.51	&	11.8	\\
2433098.67	&	12.1	\\
2437851.91	&	$>$14.2	\\
2440031.74	&	11.4	\\
2440383.73	&	$>$13.5	\\
2442164.78	&	12.8	\\
2442217.70	&	12.7	\\
2442219.69	&	12.6	\\
2442574.75	&	12.7	\\
2442624.62	&	12.0	\\
2442684.52	&	12.5	\\
2442713.49	&	12.4	\\
2442714.49	&	12.2	\\
2442849.89	&	12.5	\\
2442902.80	&	12.5	\\
2442923.80	&	12.4	\\
2442923.82	&	12.5	\\
2442932.74	&	12.5	\\
2442951.74	&	12.2	\\
2442964.72	&	12.7	\\
2442992.70	&	12.8	\\
2443010.66	&	12.5	\\
2443020.61	&	12.6	\\
2443050.59	&	12.4	\\
2443280.77	&	12.9	\\
2443289.71	&	12.1	\\
2443303.77	&	12.1	\\
2443335.67	&	12.0	\\
2443339.64	&	11.9	\\
2443347.67	&	12.2	\\
2443395.64	&	12.5	\\
2443421.55	&	12.2	\\
2443448.48	&	12.2	\\
2443613.79	&	12.0	\\
2443613.84	&	11.9	\\
2443630.80	&	12.4	\\
2443640.72	&	12.2	\\
2443659.74	&	12.1	\\
2443659.77	&	12.0	\\
2443670.74	&	11.9	\\
2443688.74	&	12.4	\\
2443690.75	&	12.3	\\
2443777.59	&	12.3	\\
2443994.76	&	$>$14.0	\\
2444024.74	&	$>$14.1	\\
2444024.77	&	$>$13.8	\\
2444045.67	&	$>$14.1	\\
2444045.72	&	$>$14.1	\\
2444099.59	&	$>$14.3	\\
2444140.56	&	$>$13.5	\\
2444402.74	&	$>$14.0	\\
2444439.68	&	$>$13.7	\\
2444465.65	&	$>$13.4	\\
2444494.57	&	$>$13.9	\\
2444514.52	&	$>$13.4	\\
2444522.51	&	$>$14.1	\\
2444522.58	&	$>$13.8	\\
2444539.48	&	$>$14.1	\\
2444732.76	&	13.0	\\
2444756.79	&	13.4	\\
2444761.74	&	13.1	\\
2444836.61	&	13.3	\\
2444898.48	&	$>$13.4	\\
2445027.89	&	$>$12.8	\\
2445089.75	&	$>$14.1	\\
2445107.77	&	$>$12.6	\\
2445139.70	&	$>$13.8	\\
2445162.64	&	$>$13.2	\\
2445195.64	&	$>$14.2	\\
2445207.61	&	$>$14.2	\\
2445223.59	&	$>$14.3	\\
2445232.58	&	$>$13.9	\\
2445253.53	&	$>$13.8	\\
2445281.49	&	$>$13.7	\\
2445442.80	&	12.6	\\
2445442.86	&	12.8	\\
2445467.77	&	12.5	\\
2445472.72	&	12.9	\\
2445494.70	&	12.7	\\
2445518.70	&	12.4	\\
2445525.66	&	12.6	\\
2445550.67	&	12.4	\\
2445581.61	&	12.5	\\
2445587.57	&	12.3	\\
2445606.55	&	12.4	\\
2445636.52	&	12.5	\\
2445665.45	&	12.3	\\
2445742.90	&	12.7	\\
2445819.80	&	$>$14.2	\\
2445846.75	&	$>$13.7	\\
2445872.68	&	$>$14.3	\\
2445883.69	&	$>$13.8	\\
2445905.70	&	$>$14.0	\\
2445930.62	&	$>$13.8	\\
2445939.61	&	$>$14.1	\\
2445960.57	&	$>$13.8	\\
2445989.51	&	$>$13.8	\\
2446114.92	&	12.9	\\
2446117.92	&	12.7	\\
2446140.89	&	12.5	\\
2446150.88	&	12.9	\\
2446173.80	&	12.4	\\
2446206.73	&	12.4	\\
2446231.74	&	12.1	\\
2446236.81	&	12.7	\\
2446260.70	&	12.0	\\
2446270.69	&	12.1	\\
2446316.62	&	12.3	\\
2446325.57	&	12.5	\\
2446326.58	&	12.3	\\
2446523.83	&	12.2	\\
2446532.84	&	12.1	\\
2446532.87	&	12.0	\\
2446533.70	&	12.4	\\
2446553.76	&	12.4	\\
2446564.76	&	12.0	\\
2446584.80	&	12.3	\\
2446612.73	&	12.2	\\
2446619.64	&	11.8	\\
2446622.81	&	12.4	\\
2446637.62	&	12.0	\\
2446668.57	&	12.5	\\
2446681.60	&	12.3	\\
2446727.49	&	12.3	\\
2446732.47	&	12.2	\\
2446735.52	&	12.5	\\
2446763.48	&	12.2	\\
2446853.89	&	12.2	\\
2446915.78	&	12.1	\\
2446946.69	&	11.8	\\
2446965.69	&	12.0	\\
2446970.72	&	12.1	\\
2446975.76	&	12.1	\\
2446999.68	&	11.7	\\
2447026.67	&	12.2	\\
2447031.61	&	11.9	\\
2447064.57	&	12.3	\\
2447082.55	&	12.2	\\
2447085.50	&	12.0	\\
2447090.53	&	11.8	\\
2447210.91	&	12.0	\\
2447238.86	&	12.1	\\
2447243.87	&	12.3	\\
2447270.81	&	12.0	\\
2447277.83	&	12.0	\\
2447294.80	&	12.1	\\
2447296.77	&	12.0	\\
2447319.73	&	12.3	\\
2447353.69	&	11.8	\\
2447469.49	&	12.0	\\
2447480.51	&	12.0	\\
2447504.46	&	12.3	\\
2447525.45	&	11.9	\\
2447596.90	&	12.1	\\
2447626.87	&	12.1	\\
2447628.75	&	12.3	\\
2447651.79	&	12.0	\\
2447689.79	&	12.0	\\
2447706.70	&	11.9	\\
2447762.59	&	11.9	
\enddata
\tablenotetext{a}{The uncertainties are $\sim$0.15 mag.}
\end{deluxetable}

\begin{deluxetable}{ccc}  
\tablecolumns{3}
\tablecaption{ROTSE-I Photometry}
\tablehead{   
  \colhead{JD} &
  \colhead{V} &
  \colhead{$\sigma_V$} 
}
\startdata
2451275.3466	&	12.42	&	0.02	\\
2451277.3501	&	12.41	&	0.01	\\
2451277.3511	&	12.43	&	0.02	\\
2451283.2551	&	12.52	&	0.02	\\
2451286.3526	&	12.60	&	0.02	\\
2451286.3537	&	12.58	&	0.02	\\
2451287.3534	&	12.58	&	0.01	\\
2451287.3544	&	12.61	&	0.02	\\
2451288.3543	&	12.58	&	0.01	\\
2451288.3553	&	12.58	&	0.01	\\
2451291.3096	&	12.54	&	0.02	\\
2451291.3106	&	12.54	&	0.02	\\
2451295.3537	&	12.46	&	0.02	\\
2451295.3547	&	12.43	&	0.02	\\
2451297.3717	&	12.39	&	0.02	\\
2451297.3727	&	12.42	&	0.02	\\
2451300.3668	&	12.28	&	0.02	\\
2451300.3678	&	12.30	&	0.02	\\
2451304.2117	&	12.25	&	0.01	\\
2451304.3632	&	12.23	&	0.01	\\
2451304.3642	&	12.24	&	0.01	\\
2451305.2125	&	12.24	&	0.01	\\
2451306.3649	&	12.23	&	0.01	\\
2451306.3659	&	12.21	&	0.01	\\
2451308.2150	&	12.18	&	0.02	\\
2451308.2160	&	12.18	&	0.02	\\
2451311.3743	&	12.20	&	0.01	\\
2451312.3647	&	12.20	&	0.01	\\
2451318.2341	&	12.17	&	0.03	\\
2451321.3776	&	12.24	&	0.01	\\
2451322.3752	&	12.22	&	0.01	\\
2451323.3634	&	12.23	&	0.01	\\
2451323.3644	&	12.27	&	0.01	\\
2451324.2246	&	12.26	&	0.02	\\
2451324.2256	&	12.28	&	0.02	\\
2451324.3714	&	12.23	&	0.01	\\
2451325.2260	&	12.25	&	0.03	\\
2451325.2263	&	12.26	&	0.03	\\
2451325.2809	&	12.34	&	0.03	\\
2451325.2812	&	12.28	&	0.03	\\
2451328.1871	&	12.27	&	0.04	\\
2451328.2430	&	12.34	&	0.03	\\
2451330.1883	&	12.39	&	0.03	\\
2451330.2451	&	12.32	&	0.03	\\
2451332.2462	&	12.34	&	0.03	\\
2451334.2310	&	12.39	&	0.01	\\
2451335.2316	&	12.42	&	0.01	\\
2451335.2326	&	12.40	&	0.01	\\
2451336.2332	&	12.37	&	0.02	\\
2451336.3892	&	12.41	&	0.01	\\
2451337.2327	&	12.47	&	0.01	\\
2451337.2337	&	12.45	&	0.01	\\
2451337.3897	&	12.46	&	0.01	\\
2451337.3907	&	12.45	&	0.01	\\
2451338.2332	&	12.45	&	0.01	\\
2451338.2342	&	12.48	&	0.02	\\
2451338.3902	&	12.47	&	0.01	\\
2451338.3912	&	12.46	&	0.01	\\
2451339.2336	&	12.49	&	0.02	\\
2451339.2347	&	12.48	&	0.02	\\
2451339.3906	&	12.48	&	0.01	\\
2451339.3916	&	12.48	&	0.01	\\
2451340.2274	&	12.50	&	0.01	\\
2451340.2284	&	12.50	&	0.01	\\
2451340.3943	&	12.50	&	0.01	\\
2451341.2325	&	12.49	&	0.01	\\
2451341.2335	&	12.52	&	0.01	\\
2451341.3894	&	12.55	&	0.02	\\
2451341.3904	&	12.52	&	0.02	\\
2451348.3089	&	12.53	&	0.02	\\
2451348.3099	&	12.53	&	0.02	\\
2451352.2334	&	12.47	&	0.02	\\
2451352.2344	&	12.47	&	0.02	\\
2451352.3862	&	12.49	&	0.01	\\
2451352.3872	&	12.48	&	0.01	\\
2451353.2829	&	12.48	&	0.02	\\
2451353.2833	&	12.46	&	0.02	\\
2451353.3387	&	12.46	&	0.02	\\
2451354.1960	&	12.47	&	0.03	\\
2451354.1964	&	12.51	&	0.03	\\
2451356.2066	&	12.38	&	0.03	\\
2451358.1957	&	12.35	&	0.03	\\
2451359.2428	&	12.37	&	0.03	\\
2451359.2431	&	12.35	&	0.03	\\
2451359.2995	&	12.37	&	0.02	\\
2451361.2545	&	12.31	&	0.03	\\
2451361.2548	&	12.31	&	0.03	\\
2451362.1948	&	12.31	&	0.02	\\
2451362.1951	&	12.30	&	0.02	\\
2451364.2336	&	12.26	&	0.02	\\
2451364.2339	&	12.31	&	0.02	\\
2451364.2920	&	12.27	&	0.02	\\
2451364.2924	&	12.26	&	0.02	\\
2451370.2307	&	12.23	&	0.01	\\
2451370.2317	&	12.25	&	0.01	\\
2451370.3866	&	12.23	&	0.02	\\
2451373.2314	&	12.22	&	0.01	\\
2451373.2324	&	12.26	&	0.01	\\
2451373.3873	&	12.25	&	0.01	\\
2451375.2303	&	12.25	&	0.01	\\
2451375.2313	&	12.24	&	0.01	\\
2451375.3784	&	12.23	&	0.01	\\
2451375.3794	&	12.23	&	0.01	\\
2451378.2265	&	12.23	&	0.01	\\
2451382.2217	&	12.19	&	0.02	\\
2451382.3797	&	12.24	&	0.01	\\
2451389.2382	&	12.27	&	0.03	\\
2451390.1794	&	12.26	&	0.02	\\
2451390.1797	&	12.26	&	0.02	\\
2451390.2375	&	12.25	&	0.02	\\
2451390.2378	&	12.27	&	0.02	\\
2451402.2079	&	12.20	&	0.01	\\
2451402.3650	&	12.20	&	0.01	\\
2451402.3660	&	12.19	&	0.01	\\
2451403.2073	&	12.19	&	0.01	\\
2451403.2083	&	12.19	&	0.01	\\
2451403.3623	&	12.18	&	0.01	\\
2451403.3633	&	12.15	&	0.01	\\
2451404.2075	&	12.15	&	0.02	\\
2451408.3525	&	12.12	&	0.01	\\
2451408.3535	&	12.10	&	0.01	\\
2451409.3985	&	12.13	&	0.01	\\
2451413.1923	&	12.09	&	0.01	\\
2451414.2447	&	12.09	&	0.02	\\
2451414.2451	&	12.06	&	0.02	\\
2451415.3234	&	12.04	&	0.03	\\
2451415.3237	&	12.11	&	0.03	\\
2451415.3783	&	12.08	&	0.04	\\
2451415.3786	&	12.07	&	0.03	\\
2451418.2478	&	12.09	&	0.03	\\
2451421.2483	&	12.16	&	0.02	\\
2451421.3044	&	12.11	&	0.02	\\
2451421.3047	&	12.15	&	0.02	\\
2451425.2525	&	12.25	&	0.02	\\
2451425.2528	&	12.19	&	0.02	\\
2451426.1777	&	12.24	&	0.01	\\
2451426.1787	&	12.25	&	0.01	\\
2451426.3306	&	12.24	&	0.01	\\
2451427.1766	&	12.29	&	0.01	\\
2451427.1776	&	12.30	&	0.01	\\
2451427.3307	&	12.22	&	0.02	\\
2451437.2363	&	12.45	&	0.02	\\
2451439.2217	&	12.49	&	0.02	\\
2451439.2227	&	12.46	&	0.02	\\
2451441.1628	&	12.49	&	0.02	\\
2451441.1638	&	12.50	&	0.02	\\
2451442.1238	&	12.51	&	0.02	\\
2451442.1242	&	12.53	&	0.02	\\
2451443.2224	&	12.52	&	0.03	\\
2451443.2787	&	12.58	&	0.03	\\
2451444.1761	&	12.53	&	0.03	\\
2451444.1764	&	12.55	&	0.03	\\
2451448.1171	&	12.50	&	0.02	\\
2451448.1174	&	12.49	&	0.02	\\
2451448.1734	&	12.48	&	0.03	\\
2451448.1737	&	12.54	&	0.03	\\
2451449.3238	&	12.39	&	0.04	\\
2451449.3241	&	12.44	&	0.04	\\
2451452.1202	&	12.35	&	0.02	\\
2451452.1206	&	12.36	&	0.02	\\
2451452.1778	&	12.37	&	0.02	\\
2451452.1781	&	12.40	&	0.02	\\
2451453.1473	&	12.37	&	0.01	\\
2451453.1483	&	12.38	&	0.01	\\
2451455.1839	&	12.32	&	0.01	\\
2451455.1850	&	12.35	&	0.01	\\
2451456.1449	&	12.35	&	0.01	\\
2451456.2872	&	12.26	&	0.02	\\
2451456.2882	&	12.34	&	0.02	\\
2451457.1439	&	12.32	&	0.01	\\
2451457.1449	&	12.31	&	0.01	\\
2451460.1410	&	12.28	&	0.01	\\
2451460.1420	&	12.29	&	0.01	\\
2451461.1422	&	12.28	&	0.01	\\
2451461.1432	&	12.30	&	0.01	\\
2451463.1402	&	12.27	&	0.01	\\
2451463.1412	&	12.27	&	0.01	\\
2451464.1394	&	12.26	&	0.01	\\
2451464.1404	&	12.27	&	0.01	\\
2451465.1384	&	12.21	&	0.01	\\
2451465.1394	&	12.23	&	0.01	\\
2451466.1375	&	12.22	&	0.01	\\
2451466.1385	&	12.23	&	0.01	\\
2451467.1366	&	12.25	&	0.01	\\
2451467.1376	&	12.23	&	0.01	\\
2451473.1370	&	12.18	&	0.02	\\
2451475.1170	&	12.22	&	0.02	\\
2451478.0896	&	12.17	&	0.02	\\
2451478.0899	&	12.17	&	0.02	\\
2451478.1469	&	12.20	&	0.02	\\
2451479.0895	&	12.17	&	0.02	\\
2451479.0899	&	12.17	&	0.02	\\
2451487.1219	&	12.24	&	0.01	\\
2451488.1225	&	12.29	&	0.01	\\
2451489.1209	&	12.29	&	0.01	\\
2451489.1219	&	12.30	&	0.01	\\
2451492.1627	&	12.31	&	0.02	\\
2451492.1637	&	12.28	&	0.02	\\
2451494.1619	&	12.34	&	0.02	\\
2451495.1624	&	12.27	&	0.02	\\
2451496.1600	&	12.22	&	0.02	\\
2451497.1566	&	12.25	&	0.02	\\
2451498.1561	&	12.28	&	0.02	\\
2451498.1571	&	12.25	&	0.02	\\
2451504.1516	&	12.28	&	0.02	\\
2451508.1178	&	12.33	&	0.03	\\
2451603.4437	&	12.34	&	0.01	\\
2451611.3877	&	12.34	&	0.01	\\
2451627.3378	&	12.15	&	0.02	\\
2451627.3382	&	12.17	&	0.03	
\enddata
\end{deluxetable}

\clearpage

\begin{deluxetable}{ccccc}  
\tablecolumns{5}
\tablecaption{SRO Photometry}
\tablehead{   
  \colhead{JD} &
  \colhead{B} &
  \colhead{V} &
  \colhead{R$_C$}&
    \colhead{I$_C$}
}
\startdata
2455228	&	13.16	&	12.04	&	11.47	&	10.85	\\
2455229	&	\nodata	&	12.07	&	\nodata	&	\nodata	\\
2455233	&	13.08	&	11.96	&	11.43	&	10.80	\\
2455237	&	13.01	&	11.92	&	11.35	&	10.76	\\
2455240	&	12.96	&	11.87	&	11.32	&	10.73	\\
2455261	&	13.14	&	12.02	&	11.55	&	10.86	\\
2455265	&	13.16	&	12.04	&	11.50	&	10.89	\\
2455266	&	13.14	&	12.04	&	11.47	&	10.87	\\
2455273	&	13.09	&	11.97	&	11.40	&	10.81	\\
2455281	&	13.02	&	11.91	&	11.39	&	10.77	\\
2455282	&	13.01	&	11.90	&	11.35	&	10.77	\\
2455286	&	12.96	&	11.87	&	11.30	&	10.73	\\
2455287	&	12.95	&	11.89	&	11.29	&	10.73	\\
2455288	&	12.96	&	11.90	&	11.29	&	10.72	\\
2455290	&	12.92	&	11.83	&	11.26	&	10.71	\\
2455291	&	12.93	&	11.83	&	11.27	&	10.71	\\
2455294	&	12.94	&	11.83	&	11.28	&	10.74	\\
2455298	&	13.01	&	11.91	&	11.33	&	10.75	\\
2455301	&	13.06	&	11.95	&	11.38	&	10.78	\\
2455304	&	13.13	&	12.01	&	11.43	&	10.81	\\
2455305	&	13.16	&	12.02	&	11.46	&	10.84	\\
2455308	&	13.18	&	12.05	&	11.46	&	10.86	\\
2455311	&	13.15	&	12.03	&	11.45	&	10.84	\\
2455312	&	13.13	&	12.01	&	11.44	&	10.84	\\
2455313	&	13.14	&	12.01	&	11.43	&	10.94	\\
2455314	&	13.16	&	12.01	&	11.42	&	10.83	\\
2455315	&	13.14	&	12.00	&	11.42	&	10.83	\\
2455316	&	13.12	&	11.99	&	11.44	&	10.83	\\
2455317	&	13.11	&	11.98	&	11.40	&	10.82	\\
2455318	&	13.11	&	11.98	&	11.40	&	10.82	\\
2455319	&	13.10	&	11.97	&	11.39	&	10.81	\\
2455320	&	13.09	&	11.96	&	11.39	&	10.81	\\
2455321	&	13.08	&	11.95	&	11.44	&	10.81	\\
2455322	&	13.08	&	11.95	&	11.45	&	10.81	\\
2455323	&	13.07	&	11.94	&	11.38	&	10.81	\\
2455327	&	13.04	&	11.92	&	11.35	&	10.77	\\
2455328	&	13.07	&	11.90	&	11.36	&	10.76	\\
2455329	&	12.99	&	11.89	&	11.33	&	10.75	\\
2455330	&	12.97	&	11.87	&	11.42	&	10.74	\\
2455333	&	12.95	&	11.86	&	11.28	&	10.71	\\
2455336	&	12.94	&	11.83	&	11.38	&	10.76	\\
2455342	&	13.01	&	11.89	&	11.53	&	10.79	\\
2455343	&	13.00	&	11.89	&	11.51	&	10.78	\\
2455344	&	13.01	&	11.91	&	11.35	&	10.77	\\
2455346	&	13.03	&	11.92	&	11.39	&	10.80	\\
2455347	&	13.05	&	11.94	&	11.45	&	10.84	\\
2455348	&	13.06	&	11.94	&	11.51	&	10.82	\\
2455349	&	13.07	&	11.95	&	11.44	&	10.83	\\
2455351	&	13.08	&	11.97	&	11.41	&	10.83	\\
2455369	&	13.06	&	11.96	&	11.43	&	10.83	\\
2455371	&	13.05	&	11.95	&	11.46	&	10.82	\\
2455372	&	13.06	&	11.94	&	11.43	&	10.81	\\
2455445	&	13.15	&	12.04	&	11.52	&	10.89	\\
2455449	&	13.18	&	12.05	&	11.49	&	10.88	\\
2455455	&	13.07	&	11.98	&	11.42	&	10.83	\\
2455456	&	13.07	&	11.96	&	11.54	&	10.86	\\
2455457	&	13.06	&	11.95	&	11.39	&	10.81	\\
2455459	&	13.05	&	11.94	&	11.38	&	10.79	\\
2455460	&	13.06	&	11.95	&	11.37	&	10.80	\\
2455465	&	13.11	&	11.99	&	11.57	&	10.87	\\
2455469	&	13.21	&	12.07	&	11.52	&	10.89	\\
2455471	&	13.26	&	12.12	&	11.56	&	10.93	\\
2455472	&	13.28	&	12.14	&	11.57	&	10.94	\\
2455477	&	13.33	&	12.20	&	11.61	&	11.00	\\
2455571	&	13.15	&	12.04	&	11.47	&	10.83	\\
2455572	&	13.17	&	12.05	&	11.49	&	10.85	\\
2455573	&	13.17	&	12.05	&	11.49	&	10.87	\\
2455577	&	13.17	&	12.04	&	11.48	&	10.83	\\
2455578	&	13.16	&	12.04	&	11.47	&	10.85	\\
2455586	&	12.99	&	11.87	&	11.31	&	10.71	\\
2455587	&	12.95	&	11.85	&	11.30	&	10.71	\\
2455588	&	12.95	&	11.83	&	11.28	&	10.68	\\
2455591	&	12.92	&	11.82	&	11.27	&	10.66	\\
2455599	&	12.96	&	11.86	&	11.30	&	10.70	\\
2455600	&	12.98	&	11.87	&	11.31	&	10.69	\\
2455611	&	13.20	&	12.03	&	11.46	&	10.84	\\
2455615	&	13.20	&	12.08	&	11.51	&	10.85	\\
2455617	&	\nodata	&	12.10	&	\nodata	&	\nodata	\\
2455625	&	13.28	&	12.15	&	11.56	&	10.89	\\
2455633	&	13.18	&	12.05	&	11.43	&	10.80	\\
2455635	&	13.09	&	11.96	&	11.40	&	10.76	\\
2455638	&	13.03	&	11.99	&	11.37	&	10.73	\\
2455656	&	13.03	&	11.92	&	11.37	&	10.77	\\
2455657	&	13.05	&	11.94	&	11.39	&	10.80	\\
2455669	&	13.11	&	12.00	&	11.44	&	10.83	\\
2455670	&	13.11	&	11.99	&	11.44	&	10.82	\\
2455672	&	13.09	&	11.96	&	11.40	&	10.79	\\
2455673	&	13.05	&	11.92	&	11.38	&	10.78	\\
2455675	&	12.97	&	11.86	&	11.30	&	10.70	\\
2455703	&	12.93	&	11.84	&	11.30	&	10.72	\\
2455704	&	12.95	&	11.87	&	11.32	&	10.73	\\
2455705	&	12.99	&	11.89	&	11.35	&	10.76	\\
2455706	&	13.02	&	11.92	&	11.37	&	10.79	\\
2455708	&	13.06	&	11.97	&	11.42	&	10.81	\\
2455710	&	13.12	&	12.00	&	11.45	&	10.85	\\
2455717	&	13.23	&	12.09	&	11.54	&	10.91	\\
2455718	&	13.23	&	12.10	&	11.55	&	10.92	\\
2455719	&	13.26	&	12.12	&	11.56	&	10.92	\\
2455720	&	13.25	&	12.12	&	11.55	&	10.91	\\
2455721	&	13.24	&	12.12	&	11.55	&	10.92	\\
2455722	&	13.26	&	12.12	&	11.56	&	10.93	\\
2455723	&	13.25	&	12.12	&	11.55	&	10.91	\\
2455724	&	13.24	&	12.11	&	11.55	&	10.92	\\
2455727	&	13.18	&	12.05	&	11.48	&	10.86	\\
2455728	&	13.16	&	12.02	&	11.45	&	10.83	\\
2455729	&	13.09	&	11.96	&	11.40	&	10.80	\\
2455730	&	13.05	&	11.93	&	11.37	&	10.76	\\
2455731	&	13.01	&	11.88	&	11.32	&	10.72	\\
2455732	&	12.98	&	11.86	&	11.29	&	10.72	\\
2455734	&	12.92	&	11.82	&	11.25	&	10.67	\\
2455739	&	12.85	&	11.75	&	11.19	&	10.67	\\
2455740	&	12.84	&	11.74	&	11.19	&	10.62	\\
2455741	&	12.84	&	11.74	&	11.19	&	10.63	
\enddata
\tablenotetext{a}{The uncertainties are $\sim$0.01-0.02 mag. }
\end{deluxetable}

\clearpage

   \begin{deluxetable}{lcc}  
\tablecaption{Photometry of NSV 11154}
\tablehead{   
  \colhead{Bands} &
  \colhead{Flux(Jy)} &
  \colhead{$\sigma$}
}
\startdata
U	&	2.50E-02	&	2.30E-04	\\
V	&	6.28E-02	&	1.16E-05	\\
R$_C$	&	8.40E-02	&	7.69E-04	\\
I$_C$	&	1.22E-01	&	1.12E-03	\\
J	&	1.45E-01	&	3.00E-03	\\
H	&	1.48E-01	&	3.00E-03	\\
K	&	1.76E-01	&	2.00E-03	\\
AKARI/9	&	6.29E-01	&	1.10E-02	\\
IRAS/12	&	5.02E-01	&	2.50E-02	\\
AKARI/18	&	2.98E-01	&	3.10E-02	\\
IRAS/25	&	2.20E-01	&	1.98E-02
\enddata
\end{deluxetable}

\begin{deluxetable}{ccc}  
\tablecolumns{2}
\tablecaption{Decline Epochs of NSV 11154}
\tablehead{   
  \colhead{JD$^a$} &
  \colhead{Length (d)} 
  }
\startdata
2415150	&	\nodata	\\
2419306	&	\nodata	\\
2423981	&	$\sim$2525	\\
2427246	&	\nodata	\\
2428314	&	\nodata	\\
2430616	&	\nodata	\\
2437851	&	\nodata	\\
2440150	&	$\sim$1775	\\
2444099	&	$\sim$375	\\
2445027	&	$>$150	\\
2445742	&	$\sim$400	\\
2448682	&	$\sim$375	\\
2449422	&	$>$450
\enddata
\tablenotetext{a}{The photometry shown in Figures 2 and 3 is fragmentary
 so the decline epochs and lengths of the declines are best estimates.}
\end{deluxetable}

\end{document}